\documentclass[10pt,english]{article}
\usepackage[T1]{fontenc}
\usepackage[latin1]{inputenc}
\usepackage{graphicx}
\usepackage{amssymb}

\makeatletter

\usepackage{helvet}

\usepackage{geometry}



\usepackage{babel}
\makeatother
\begin{document}

\smallskip{}
\title{Entanglement and Fidelity Signatures of Quantum Phase Transitions
in Spin Liquid Models}

\author{Amit Tribedi and Indrani Bose }

\maketitle
\begin{center}Department of Physics \end{center}

\begin{center}Bose Institute \end{center}

\begin{center}93/1, Acharya Prafulla Chandra Road \end{center}

\begin{center}Kolkata - 700 009, India \end{center}

\begin{abstract}
We consider a spin ladder model which is known to have matrix product
states as exact ground states with spin liquid characteristics. The
model has two critical-point transitions at the parameter values $u=0$
and $\infty$. We study the variation of entanglement and fidelity
measures in the ground states as a function of $u$ and specially
look for signatures of quantum phase transitions at $u=0$ and $\infty$.
The two different entanglement measures used are $S(i)$ (the single-site
von Neumann entropy) and $S(i,j)$ (the two-body entanglement). At
the quantum critical point (QCP) $u=\infty$, the entanglement measure
$E$ $[=S(i),\: S(i,j)]$ vanishes but remains non-zero at the other
QCP $u=0$. The first and second derivative of $E$ with respect to
the parameter $u$ and the entanglement length associated with $S(i,j)$
are further calculated to identify special features, if any, near
the QCPs. We further determine the GS fidelity $F$ and a quantity
$ln|D|$ related to the second derivative of $F$ and show that these
quantities calculated for finite-sized systems are good indicators
of QPTs occurring in the infinite system. 
\end{abstract}

\section*{I. INTRODUCTION}

\smallskip{}
In recent years, quantum phase transitions (QPTs) in many-particle
systems have been extensively investigated using well-known quantum
information theoretic measures. QPTs which are solely driven by quantum
fluctuations occur at zero temperature when some parameter, either
external or intrinsic to the Hamiltonian, is tuned to a special value
termed the transition point \cite{key-1}. In the case of second-order
QPTs (critical-point transitions), a diverging length scale governs
the physics near a quantum critical point (QCP). Usually, the correlation
length associated with specific correlation functions diverges as
the QCP is approached and the ground state properties develop non-analytic
features. In this context, it is pertinent to ask whether the quantum
correlations associated with entanglement are good indicators of QPTs.
A number of entanglement measures have so far been identified which
show special features close to the transition points of QPTs occuring
in spin systems \cite{key-2,key-3,key-4,key-5,key-6,key-7}. It has
been shown that, in general, a first order QPT, linked to a discontinuity
in the first derivative of the ground state energy, is signalled by
a discontinuity in a bipartite entanglement measure such as negativity
and concurrence \cite{key-8,key-9,key-10} and a discontinuity or
a divergence in the first derivative of the same entanglement measure
marks a second order phase transition characterized by a discontinuity
or a divergence in the second derivative of the ground state energy
\cite{key-2,key-3}. The entropy of entanglement of a block of $L$
contiguous spins in a chain with the rest of the system has been shown
to diverge logarithmically with $L$ near the QCP \cite{key-4}.

The typical length scale over which a particular entanglement measure
decays defines the entanglement length (EL). A number of entanglement
measures characterized by a diverging EL close to a QCP have been
proposed to date \cite{key-7,key-11}. One of these, the two-body
entanglement $S(i,j)$ which estimates the amount of non-local correlations
between a pair of separated spins at sites $i$ and $j$ and the rest
of the spins, is given by the von Neumann entropy

\begin{equation}
S(i,j)=-Tr\,\rho(i,j)\, log_{2}\,\rho(i,j)\label{1}\end{equation}
where $\rho(i,j)$ is the two-site reduced density matrix obtained
by tracing out the spins except the ones at sites $i$ and $j$ from
the full density matrix. When the system is translationally invariant,
$S$ depends only on the separation $n=\mid j-i\mid$ and can be expressed
in terms of the spin correlation functions in the large $n$ limit.
Away from the critical point, $S(i,j)$ saturates over a length scale
$\xi_{E}$, the EL, as $n$ increases. Close to the QCP and for large
$n$, we have

\begin{equation}
S(n)-S(\infty)\sim A(n)\, e^{-\frac{n}{\xi_{E}}}\label{2}\end{equation}
$S(i,j)$ captures the long-range correlations at a QCP if $A(n)$
has a power-law decay as a function of $n$ along with a divergent
$\xi_{E}$. This is true for spin models such as the $S=\frac{1}{2}$
exactly solvable anisotropic XY model in a transverse magnetic field
\cite{key-7}. The EL is found to diverge with the same critical exponent
as the correlation length at the QCP. $S(i,j)$ and its first derivative
have been found to develop special features in the vicinity of the
QCP \cite{key-7,key-18,key-19,key-20}. The single-site von Neumann
entropy (a measure of the entanglement of a single spin with the rest
of the system)

\begin{equation}
S(i)=-Tr\,\rho(i)\, log_{2}\,\rho(i)\label{3}\end{equation}
is also known to be a good indicator of QPTs \cite{key-3,key-18,key-19}.
In Eq. (3), $\rho(i)$ is the single-site reduced density matrix.

The exploration of the entanglement properties of the ground state
of a number of spin$-1$ Hamiltonians (the AKLT model is an example)
has been carried out using both analytical and numerical techniques
\cite{key-11,key-12,key-13}. Certain spin-$1$ and generalized spin-$\frac{1}{2}$
ladder model systems are known to have matrix product (MP) states
as exact ground states \cite{key-14,key-15,key-16} . The MP states
are finitely correlated states with short-ranged spin-spin correlations,
may have hidden topological order and have gapped excitation spectra.
The second order transitions in these so-called finitely correlated
MP states belong to the class of generalized QPTs (the definition
encompasses the transitions marked by a non-analyticity in any observable
of the system) \cite{key-16} which differ from the conventional QPTs
in some important aspects. The spin correlation function in both the
cases is of the form $A_{C}\, e^{-\frac{n}{\xi_{C}}}$ for large $n$.
In the case of MP states, $A_{C}$ vanishes at the transition point
though the correlation length $\xi_{C}$ blows up as the transition
point is approached. In the case of a conventional QCP, the correlation
function has a power-law decay close to the QCP. A distinguishing
feature of QPTs in MP states arises from the fact that the ground
state energy density is analytic for all values of the control parameter.
A critical point transition is, however, still signalled by a diverging
correlation length and the vanishing of an energy gap. The MP states
have been used as trial wave functions for a number of standard spin
models and provide the basis for the well-known density matrix renormalization
group (DMRG) method leading to several interesting developments in
quantum information theory \cite{key-17}. The MP states further serve
as candidate systems for the study of unconventional QPTs.

Recently, ground state fidelity has been proposed to provide a signature
of QPTs \cite{key-21,key-22} and the usefulness of the measure has
been explored in a number of studies \cite{key-23,key-24,key-25,key-26,key-27,key-28,key-29}.
Fidelity, a concept borrowed from quantum information theory, is defined
as the overlap modulus between ground states corresponding to slightly
different Hamiltonian parameters. The advantage of using this measure
is that it characterizes QPTs without needing any $a\, priori$ knowledge
of the order parameter and the symmetries of the system. The fidelity
typically drops in an abrupt manner at a transition point indicating
a dramatic change in the nature of the ground state wave function.
A QCP is characterized by the vanishing of the single particle excitation
gap. In Ref. \cite{key-27}, an explicit connection between the vanishing
of the gap and the fidelity drop has been established. Cozzini et
al. \cite{key-23} tested the validity of the fidelity approach for
probing QPTs in MP states and also studied the finite size scaling
of the fidelity derivative establishing its relevance in extracting
critical exponents. The QPT in the Bose-Hubbard model which is difficult
to detect using conventional entanglement measures has been correctly
predicted using the fidelity measure \cite{key-25}. Chen et al. \cite{key-29}
have shown that the fidelity of the first excited state and not the
ground state, is the appropriate quantity to signal QPTs in models
such as the antiferromagnetic (AFM) Heisenberg spin chain with nearest-neighbour
as well as next-nearest-neighbour interactions.

In this paper, we study a $S=\frac{1}{2}$ ladder model with MP states
as exact ground states \cite{key-30}. The model has an interesting
phase diagram with two critical point transitions. We explore the
properties of the ground state using two different entanglement measures,
namely, the single-site and the two-body entanglement. The major motivation
is to identify distinctive features, if any, in the entanglement measures
close to the QCPs. We look at the same QPTs in the light of fidelity
analysis and show that the fidelity $\mathcal{F}$ of the ground state
is an efficient indicator of the QPTs. The quantity $ln\left|D(u)\right|$,
related to the second derivative of $\mathcal{F}$ , also yields useful
information regarding the QPTs. We apply the idea of average entanglement
\cite{key-31} to take care of the two-fold degeneracy of the ground
state of the model.

\section*{II. ENTANGLEMENT MEASURES }

We consider a general ladder model proposed by Kolezhuk et al. \cite{key-30}
which is described by a Hamiltonian of the general form

\[
H=\sum_{j=1}[J(S_{1,j}S_{1,j+1}+S_{2,j}S_{2,j+1})+J_{r}S_{1,j}S_{2,j}+V(S_{1,j}S_{1,j+1})(S_{2,j}S_{2,j+1})\]

\begin{equation}
+J_{d}(S_{1,j}S_{2,j+1}+S_{2,j}S_{1,j+1})+K\{(S_{1,j}S_{2,j+1})(S_{2,j}S_{1,j+1})-(S_{1,j}S_{2,j})(S_{1,j+1}S_{2,j+1})]\label{4}\end{equation}
 where the indices $1$ and $2$ distinguish the lower and upper legs
of the ladder and $i$ labels the rungs. The ground state $\left|\psi_{0}\,(u,\,\tilde{u})\right\rangle $
has the following MP form\begin{equation}
\left|\psi_{0}\,(u,\,\tilde{u})\right\rangle =\frac{1}{\sqrt{N_{C}}}Tr\,\{ g_{1}(u).g_{2}(\tilde{u})......g_{2N-1}(u).g_{2N}(\tilde{u}))\label{5}\end{equation}
 where \begin{equation}
g_{i}(u)=\left(\begin{array}{cc}
u\left|s\right\rangle _{i}+\left|t_{0}\right\rangle _{i} & -\sqrt{2}\left|t_{+}\right\rangle _{i}\\
\sqrt{2}\left|t_{-}\right\rangle _{i} & u\left|s\right\rangle _{i}-\left|t_{0}\right\rangle _{i}\end{array}\right)\label{6}\end{equation}
and $N_{C}\:(=(u\tilde{u}+3)^{2N}+3\,(u\tilde{u}-1)^{2N})$ is the
normalization factor. Here $\left|s\right\rangle _{i}$ is the singlet
state and $\left|t_{\mu}\right\rangle $ with $\mu=+1$, $0$ and
$-1$ are the triplet states of the $i$-th rung with $S^{z}=+1$,
$0$ and $-1$, respectively. $2N$ is the total number of rungs (with
periodic boundary conditions) and $u$, $\tilde{u}$ are free parameters.
For $u\neq\tilde{u}$, $\left|\psi_{0}\,(u,\,\tilde{u})\right\rangle $
is dimerized and doubly degenerate as the translation of the rungs
by one unit leads to a different state with the same energy.

It is convenient to rewrite the Hamiltonian $(5)$ as a sum of identical
local terms which couple only neighbouring rungs, $H=\sum_{j}(h_{i,i+1}-E_{0})$.
The value of $E_{0}$ is adjusted to make $\left|\psi_{0}\right\rangle $
a zero-energy ground state which requires the following conditions
to be satisfied. (i) All elements of the two matrix products $g_{i}(u).g_{i+1}(\tilde{u})$
and $g_{i}(\tilde{u}).g_{i+1}(u)$ have to be zero-energy eigenstates
of $h_{i,i+1}$. (ii) The other eigenstates of $h_{i,i+1}$ should
have positive energy. The two conditions are satisfied when $h_{i,i+1}$
has the structure\begin{equation}
h_{i,i+1}=\sum_{J=0,1,2}\sum_{M=-J}^{J}\epsilon_{J}\left|\psi_{JM}\right\rangle \left\langle \psi_{JM}\right|\label{7}\end{equation}
where the eigenvalues $\epsilon_{J}>0$ and $\left|\psi_{JM}\right\rangle $'s
are the components of the positive-energy multiplets constructed from
the states of the two-rung plaquette $(i,\: i+1)$:\begin{equation}
\begin{array}{c}
\left|\psi_{00}\right\rangle =[3+(u\tilde{u})^{2}]^{-\frac{1}{2}}\{\sqrt{3}\left|ss\right\rangle +u\tilde{u}\left|tt\right\rangle _{J=0}\}\\
\left|\psi_{1}\right\rangle =[2+(f)^{2}]^{-\frac{1}{2}}\{\left|st\right\rangle +\left|ts\right\rangle +f\,\left|tt\right\rangle _{J=1}\}\\
\left|\psi_{1}\right\rangle =\left|tt\right\rangle _{J=2},\quad f=\frac{u\tilde{+u}}{\sqrt{2}}\end{array}\label{8}\end{equation}
The notation $\left|tt\right\rangle _{J=1}$ has been used to describe
states with the total spin $J=1$ constructed from two triplets on
rungs $i$ and $i+1$, etc. We obtain the connections between the
parameters $J$, $J_{r}$, $J_{d}$, $V$ and $K$ of Eq. $(4)$,
the local eigenvalues $\epsilon_{J}$ and the singlet weight parameters
$u,\,\tilde{u}$ of the ground state wave function by claiming that
the structure $(7)$ is compatible with Eq. $(4)$. The model we study
in this paper is a special case of the three types of solutions obtained
from the above-mentioned relationships. In this case, $J_{d}=0$,
$K\neq0$ and\begin{equation}
\begin{array}{c}
u=-\tilde{u},\: K=J_{r}=\epsilon_{0}\frac{(u^{2}-1)(u^{2}+3)}{2},\: J_{d}=0,\\
V=\epsilon_{0}\frac{(5u^{4}+2u^{2}+9)}{4},\: J=3\epsilon_{0}\frac{(u^{4}+10u^{2}+5)}{16},\\
\epsilon_{1}=\epsilon_{0}\frac{(3u^{4}+14u^{2}+15)}{8},\:\epsilon_{2}=\epsilon_{0}\frac{(5u^{4}+18u^{2}+9)}{8}\end{array}\label{9}\end{equation}

As pointed out in Ref. \cite{key-30}, the one-parameter model undergoes
two second-order phase transitions, one at $u=0$ and the other at
$u=\infty$. At $u=0$, the ground state undergoes a transition from
the dimerized phase to the Haldane phase. The effective Hamiltonian
describing this phase is that of the $S=1$ AKLT chain. At $u=\infty$,
the transition is to a phase in which the ground state is a product
of singlet bonds on the rungs. The transitions at $u=0$ and $\infty$
are marked by the vanishing of the singlet and triplet gaps, respectively,
in the excitation spectrum \cite{key-30}. The ground state is spontaneously
dimerized everywhere except at the critical points. In the MP formalism,
it is straightforward to calculate the spin-spin and dimer-dimer correlation
functions $C_{S}(n)=\left\langle S_{1,i}^{z}\, S_{1,i+n}^{z}\right\rangle $and
$C_{D}(n)=\left\langle D_{i}\, D_{i+n}\right\rangle $ where $D_{i}=\mathbf{S}_{1,i}.(\mathbf{S}_{1,i+1}-\mathbf{S}_{1,i-1})$.
The dimer correlations are long-ranged and vanish as $u\rightarrow0,\:\infty$
but with no exponential tail. The spin correlation length is finite
at the AKLT point $u=0$, becomes zero at $u=1$ and diverges as $u\rightarrow\infty$.
There is, however, no development of long-range spin order since the
amplitude of the spin correlations becomes zero in this limit. The
doubly-degenerate spontaneously dimerized phase which prevails away
from the critical points exhibits non-Haldane spin liquid properties.
The elementary excitation is of a novel type, a pair of propagating
triplet or singlet solitons connecting two spontaneously dimerized
ground states \cite{key-30}. In the Haldane phase, the elementary
excitation has the character of a magnon.

Using the transfer matrix (TM) method, we now study the entanglement
properties of the MP ground state {[}Eq. $(9)${]}. The state is two-fold
degenerate as the ground-state energy per rung $E_{0}=-\frac{3}{64}\,\lambda_{0}(7u^{4}+22u^{2}+19)$
does not depend on the sign of $u$. The two ground states obtained
from Eq. $(6)$\begin{equation}
\begin{array}{c}
\left|\psi_{1}\right\rangle =\frac{1}{\sqrt{N_{0}(u)}}Tr\,\{ g_{1}(u).g_{2}(-u)......g_{2N-1}(u).g_{2N}(-u)\}\\
\left|\psi_{2}\right\rangle =\frac{1}{\sqrt{N_{0}(u)}}Tr\,\{ g_{1}(-u).g_{2}(u)......g_{2N-1}(-u).g_{2N}(u)\}\end{array}\label{10}\end{equation}
 are asymptotically orthogonal in the thermodynamic limit (TDL) $N\rightarrow\infty$,
i.e., the overlap $\left\langle \psi_{1}|\psi_{2}\right\rangle =\frac{3\,(u^{2}+1)^{2N}+(u^{2}-3)^{2N}}{(u^{2}+3)^{2N}+3\,(u^{2}-1)^{2N}}\leq1$
for finite $N$ and vanishes in the limit $N\rightarrow\infty$. $N_{0}(u)\:[=(u^{2}+3)^{2N}+3\,(u^{2}-1)^{2N}]$
is the normalization factor. We construct a pair of orthogonal degenerate
ground states applying the usual Gram-Schmidt procedure \begin{equation}
\begin{array}{c}
\left|\phi_{1}\right\rangle =\left|\psi_{1}\right\rangle \\
\left|\phi_{2}\right\rangle =\frac{1}{\sqrt{\widetilde{N}}}(\left|\psi_{2}\right\rangle -\left\langle \psi_{1}|\psi_{2}\right\rangle \left|\psi_{1}\right\rangle )\end{array}\label{11}\end{equation}
 with $\widetilde{N}=1-\left|\left\langle \psi_{1}|\psi_{2}\right\rangle \right|^{2}$.
An arbitrary superposition of the two degenerate ground states is
also a valid ground state. We apply the idea of average entanglement
\cite{key-31}, i.e., calculate the entanglement content of a general
state (an arbitrary superposition of basis states) and then calculate
its average value over the whole of parameter space (the coefficients
of the basis-state expansion constitute the parameters) \begin{equation}
E_{av}=\frac{\int d\mu(p_{1},p_{2},...)\left|E(p_{1},p_{2},...)\right|}{\int d\mu(p_{1},p_{2},...)}\label{12}\end{equation}
 where $\int d\mu(p_{1},p_{2},...)$ is the Haar measure associated
with the parametrization $p_{1},p_{2},...,$ which is invariant under
unitary operations. The normalization condition restricts the values
of the coefficients so that the parameter space is associated with
a compact hyper-surface. In the case of a double degenerate ground
state, a general state is a superposition of two states \begin{equation}
\begin{array}{c}
\left|\phi_{s}\right\rangle =a\left|\phi_{1}\right\rangle +b\left|\phi_{2}\right\rangle \end{array}\label{13}\end{equation}
with the restriction $|a|^{2}+|b|^{2}=1$. The corresponding parameter
space is a $3-D$ sphere $S^{3}$. The one-rung reduced density matrix
$\rho(i)$ (Eq. $(3)$) is obtained by tracing out all the rungs except
the $i$-th one from the ground state density matrix $\rho=\left|\phi_{s}\right\rangle $$\left\langle \phi_{s}\right|$
. From Eq. $(13)$ \begin{equation}
\rho(i)=Tr_{1,..L}^{i}\left|\phi_{s}\right\rangle \left\langle \phi_{s}\right|=Tr_{1,..L}^{i}(|a|^{2}\left|\phi_{1}\right\rangle \left\langle \phi_{1}\right|+|b|^{2}\left|\phi_{2}\right\rangle \left\langle \phi_{2}\right|+ab^{*}\left|\phi_{1}\right\rangle \left\langle \phi_{2}\right|+a^{*}b\left|\phi_{2}\right\rangle \left\langle \phi_{1}\right|\label{14}\end{equation}
With the help of standard TM calculations \cite{key-19} , one obtains
a form for $\rho(i)$ which is found to be independent of the parameters
$a$ and $b$ in the TDL, \begin{equation}
\rho(i)=\left(\begin{array}{cccc}
\frac{1}{u^{2}+3} & 0 & 0 & 0\\
0 & \frac{1}{u^{2}+3} & 0 & 0\\
0 & 0 & \frac{1}{u^{2}+3} & 0\\
0 & 0 & 0 & \frac{u^{2}}{u^{2}+3}\end{array}\right)\label{15}\end{equation}
 in the $\left|t_{\pm1,0},s\right\rangle $ basis. The single-rung
entanglement is obtained as\begin{equation}
S(i)=\frac{1}{u^{2}+3}[(u^{2}+3)log_{2}\,(u^{2}+3)-u^{2}\, log_{2}\, u^{2}]\label{16}\end{equation}
Entanglement average, as defined in Eq. $(12)$, is required for finite-sized
systems. In the TDL, such averaging is not necessary as $\rho(i)$
{[}Eq. $(14)${]} is independent of $a$ and $b$ ($|a|^{2}+|b|^{2}=1$).
The variations of $S(i)$ and its first derivative with respect to
$u$ have been shown in Fig. $1$ $(top)$ and $(bottom)$ respectively.
$S(i)$ has the value $log_{2}\,3$ at the critical point $u=0$ (the
AKLT point) as expected, increases as $u$ is increased from zero
before it reaches its maximum possible value of $2$ at $u=1$. Then
it decreases with increasing $u$ and vanishes at the other transition
point $u=\infty$ (Fig. $2$). In the rung-singlet phase, each pair
of spins in a rung forms a singlet to become maximally entangled with
each other and completely unentangled with the rest of the system.
The plots are expectedly symmetric about the point $u=0$.

The two-rung reduced density matrix $\rho(i,j)$ can be calculated
in the same manner. $\rho(i,j)$ is given by\begin{equation}
\rho(i,j)=Tr_{1,..L}^{i,j}\left|\phi_{s}\right\rangle \left\langle \phi_{s}\right|\label{17}\end{equation}
 where the trace is taken over all the rungs except the $i$-th and
$j$-th ones. From the usual TM calculations , we obtain $\rho(i,j)$,
in the TDL, as a $16\times16$ matrix in block-diagonal form. From
$(1)$ and $(17)$, the two-body entanglement is \begin{equation}
S(i,j)=-\sum_{i}\lambda_{i}\, log_{2}\,\lambda_{i}\label{18}\end{equation}
 $\lambda_{i}$'s being the eigenvalues of $\rho(i,j)$. Figure $3$
shows the variation of the average $S(i,j)$ (top) and its first derivative
(bottom) with $u$ for $n=1000$. $S(i,j)$ behaves in a similar manner
as $S(i)$. It has the value $2\, log_{2}3$ at the QCP $u=0$, it
then increases with $u$ to attain the peak value $4$ at $u=1$ and
when $u$ is increased further, $S(i,j)$ decreases and falls to zero
(Fig. $4$) as we approach the QCP $u=\infty$ . The first derivatives
of $S(i)$ and $S(i,j)$, instead of showing any non-analyticity,
fall sharply to zero at both the QCPs. The first derivatives are also
zero at $u=1$ where the entanglement measures have the maximum value.
The second derivatives of $S(i)$ and $S(i,j)$ are logarithmically
divergent at both the QCPs $u=0$ and $\infty$ (as can be seen in
the insets of Figs. $1-4$). Both the measures $S(i)$ and $S(i,j)$
vanish at $u=\infty$ and are non-zero elsewhere, they thus behave
as an order parameter decreasing to zero value at the QCP $u=\infty$
with transition to the rung-singlet phase. The measures, however,
do not have the character of an order parameter for the transition
at $u=0$ from the dimerized to the Haldane phase.

We next calculate the EL, $\xi_{E}$, associated with the entanglement
measure $S(i,j)$ . Close to either of the QCPs and in the limit of
large $n$, we have $S(n=|j-i|)-S(\infty)\sim A_{e}\, e^{-\frac{n}{\xi_{E}}}$
. The pre-factor $A_{e}$ is found to remain finite and non-zero at
the transition point $u=0$ but it vanishes at $u=\infty$ . The EL
$\xi_{E}$ is given by\begin{equation}
\xi_{E}=\frac{1}{2\, ln|\frac{u^{2}+3}{u^{2}-1}|}\label{19}\end{equation}
 We rewrite $\xi_{E}$ as a function of $\frac{1}{u}$, i.e., $\xi_{E}=\frac{1}{2\, ln|\frac{1+3(\frac{1}{u})^{2}}{1-(\frac{1}{u})^{2}}|}$
and study its behaviour near $u=\infty$, i.e., $\frac{1}{u}=0$.
Fig. $5$ shows the variation of $\xi_{E}$ with respect to $u$ and
$\frac{1}{u}$. We find that $\xi_{E}$ is finite at the critical
point $u=0$ but it diverges as $u\rightarrow\infty$ with the critical
exponent $\nu=2$ as $\xi_{E}\sim\left(\frac{1}{u}\right)^{-2}$ for
$\frac{1}{u}\sim0$. The spin-spin correlation function $C_{S}(n)=<S_{1,i}^{z}\: S_{1,i+n}^{z}>$
can be calculated in the TM formalism as \cite{key-30} \begin{equation}
\begin{array}{c}
C_{S}(n)=(u^{2}+3)^{-1}(z_{+}z_{-})^{n}\:(\delta_{n,2k}-z_{-}\,\delta_{n,2k+1})\\
z_{\pm}=(u\pm1)^{2}/(u^{2}+3)\end{array}\label{20}\end{equation}
Close to the QCP $u=\infty$, $\xi_{E}\sim\xi_{C}/2$ so that both
$\xi_{E}$ and $\xi_{C}$ diverge with the same exponent $\nu=2$.

\section*{II. GROUND STATE FIDELITY $\mathcal{F}(u,\:\delta)$}

We now investigate the behaviour of fidelity near the same pair of
QCPs. The average fidelity, in analogy to $(12)$, is \begin{equation}
\mathcal{F}_{av}=\frac{\int d\mu(p_{1},p_{2},...)\:\mathcal{F}(p_{1},p_{2},...)}{d\mu(p_{1},p_{2},...)}\label{21}\end{equation}
The overlap between two general ground states, $\left|\phi(u_{1})\right\rangle $
and $\left|\phi(u_{2})\right\rangle $ (see Eq.$(13)$), corresponding
to two different values of the control parameter is given by\begin{equation}
\mathcal{F}(u_{1},u_{2})=\left\langle \phi(u_{1})\right|\left.\phi(u_{2})\right\rangle =|a|^{2}\left\langle \phi_{1}\right|\left.\phi_{1}\right\rangle +|b|^{2}\left\langle \phi_{2}\right|\left.\phi_{2}\right\rangle +ab^{*}\left\langle \phi_{1}\right|\left.\phi_{2}\right\rangle +a^{*}b\left\langle \phi_{2}\right|\left.\phi_{1}\right\rangle \label{22}\end{equation}
 $\mathcal{F}(u_{1},u_{2})$ (averaged over the $\{ a,b\}$ ) can
be expressed in terms of the eigenvalues of the TM \cite{key-23}
as\begin{equation}
\begin{array}{c}
\mathcal{F}(u_{1},u_{2})=\frac{1}{\sqrt{N_{0}(u_{1})N_{0}(u_{2})}}[(1+\frac{1+p(u_{1})p(u_{2})}{\sqrt{(1-p^{2}(u_{1})p^{2}(u_{2}))}})\{(u_{1}u_{2}+3)^{2N}+3\,\\
(u_{1}u_{2}-1)^{2N}\}-\frac{p(u_{1})p(u_{2})}{\sqrt{(1-p^{2}(u_{1})p^{2}(u_{2}))}}\{(u_{1}u_{2}-3)^{2N}+3\,(u_{1}u_{2}+1)^{2N}\}]\end{array}\label{23}\end{equation}
 where $p(u)=\frac{3\,(u^{2}+1)^{2N}+(u^{2}-3)^{2N}}{(u^{2}+3)^{2N}+\,3(u^{2}-1)^{2N}}$.
Fig. $6$ $(top)$ shows the nature of the variation of $\mathcal{F}(u,\: u+\delta)$,
(overlap of the states corresponding to two close points in the control
parameter space separated by a small variation) with $u$ and $N$
in the neighbourhood of the critical point $u=0$ for $\delta=.001$.
A straightforward calculation reveals that for large values of $N$
and for $u\neq0$, $\mathcal{F}(u_{1},u_{2})\thicksim(\alpha(u_{1},\, u_{2}))^{N}$,
where $\alpha(u_{1},\, u_{2})=\frac{u_{1}^{2}u_{2}^{2}+6\, u_{1}u_{2}+9}{u_{1}^{2}u_{2}^{2}+3\,(u_{1}^{2}+u_{2}^{2})+9}$.
$\alpha(u,\, u+\delta)<1$ and it has a sharp dip at $u=0$. Thus
away from the critical point, $\mathcal{F}(u,\: u+\delta)$ decreases
exponentially with $N$ and vanishes in the TDL for any fixed value
of $u$ and $\delta$, but we observe from Fig. $6$ (top) that $\mathcal{F}(u,\: u+\delta)$
decreases at a much enhanced rate when the QCP is approached. Intuitively,
the rate of orthogonality, i.e., the rate at which the {}``distance''
between the ground states corresponding to two neighbouring points
of the parameter space becomes maximal, should diverge in the proximity
of a QPT. It is thus sensible to relate the degree of criticality
to the derivative of the fidelity function. $\textrm{Cozzini et al}$
\cite{key-23} have proposed a general expression for the quantity
relevant in this case \begin{equation}
D(u)=-\partial_{u_{1}}\,\partial_{u_{2}}ln\, F(u_{1},u_{2})\left|_{u_{1}=u_{2}=u}\right.\label{24}\end{equation}
where $F(u_{1},u_{2})=\sqrt{N_{0}(u_{1})N_{0}(u_{2})}\mathcal{F}(u_{1},u_{2})$.
In the large $N$ limit and for $u\neq0$, one can easily check that
$D(u)\sim\frac{N}{(u^{2}+3)^{2}}$. Thus in the plots (Fig. $6$ (bottom)
) showing the variation of $ln|D(u)|$ with $u$ for different values
of $N$, we observe that the rate at which $ln|D(u)|$ increases with
$u$ is heightened in the proximity of the QCP $u=0$. To repeat the
whole analysis for the other critical point $u=\infty$, we express
$F(u_{1},u_{2})$ as a function of $\tilde{u_{1}}=\frac{1}{u_{1}}$
and $\tilde{u_{2}}=\frac{1}{u_{2}}$. For very large $N$, $\mathcal{F}^{'}(\tilde{u}_{1},\tilde{u}_{2})\thicksim(\alpha^{'}(u_{1},\, u_{2}))^{N}$
$[\alpha^{'}(\tilde{u}_{1},\,\tilde{u}_{2})=\frac{9\,\tilde{u}_{1}^{2}\tilde{u}_{2}^{2}+6\,\tilde{u}_{1}\tilde{u}_{2}+1}{9\,\tilde{u}_{1}^{2}\tilde{u}_{2}^{2}+3(\tilde{u}_{1}^{2}+\tilde{u}_{2}^{2})+1}]$
and $D^{'}(\tilde{u})\sim\frac{N}{(3\,\tilde{u}^{2}+1)^{2}}$ away
from the critical point. We find a similar variation of $\mathcal{F}^{'}(\tilde{u},\:\tilde{u}+\delta)$
{[}Fig. $7$ (top) {]} and $D^{'}(\tilde{u})$ {[}Fig. $7$ (bottom)
{]} near the QCP $\tilde{u}=0$, i.e., $u=\infty$ as in the case
of the QCP $u=0$. $\mathcal{F}^{'}(\tilde{u},\:\tilde{u}+\delta)$
falls sharply at $\tilde{u}=0$ and the fall becomes faster as we
increase $N$. The quantity $ln|D^{'}(\tilde{u})|$ increases at an
enhanced rate and tends to blow up in the vicinity of $\tilde{u}=0$
as we increase the value of $N$. The inset of the figure shows that
curves plotted in rescaled units collapse onto a single curve for
different values of $N$. The rescaled quantity $\frac{D^{'}(\tilde{u})}{N}$
is found to be a function of $N\tilde{u}^{2}$ only. This feature
of data collapse is analogous to the scaling behaviour of observables
in the vicinity of a critical point. The finite size scaling hypothesis,
valid in the critical region, is given by $X_{N}=N^{\frac{\rho}{\nu}}\: Q(N\,|g-g_{c}|^{\nu})$
where $X_{N}$ is some observable with the divergent behaviour $X_{N}\sim|g-g_{c}|^{-\rho}$
close to the critical point $g=g_{c}$. The exponent $\nu$ is the
correlation length exponent. In the present case $\rho\sim0$ and
$\nu=2$.

\section*{IV. DISCUSSIONS}

In this paper, we have studied a $S=\frac{1}{2}$ spin ladder model
the exact ground states of which are MP states. The ground state is
spontaneously dimerized and doubly degenerate (broken translational
symmetry) at all values of the parameter $u$ excepting the points
at $u=0$ and $u=\infty$. At $u=0$, a QPT occurs to the Haldane
phase of an effective $S=1$ chain which is signalled by the vanishing
of a singlet excitation gap. The elementary singlet excitation in
the dimerized phase is neither a magnon nor a spinon but a soliton
in the dimer order. The lowest soliton excitations occur in pairs.
At $u=\infty$, there is another QPT to the rung-singlet phase with
the vanishing of a triplet excitation gap, associated with triplet
solitons. The ground states in all the three phases: Haldane ($u=0$),
spontaneously dimerized ($0<u<\infty$) and rung-singlet ($u=\infty$)
are spin liquids with no conventional long-range order in the two-spin
correlation functions but are characterized by other types of order
parameters. The spontaneously dimerized phase has long range order
in dimer correlations which vanishes for $u\rightarrow0,\,\infty$
but there is no exponential tail. The Haldane phase has the string
order parameter \cite{key-11,key-12,key-13,key-14} whereas the rung-singlet
phase has dimer-dimer correlations with the dimers located on the
rungs. The two-spin correlation length is finite at $u=0$ and diverges
as $u\rightarrow\infty$ but no long range order develops in the latter
case since the amplitude of spin correlations falls to zero in this
limit.

As pointed out in \cite{key-16}, QPTs in MP states are unconventional
with the ground state energy analytic at $g=g_{c}$, the transition
point. A conventional QPT is signalled by a non-analyticity in the
ground state energy. One can, however, generalize the definition of
QPT to include cases where any observable quantity becomes non-analytic
as the transition point is reached. MP states are an important class
of states which provide an exact representation of many-body ground
states of specific Hamiltonians. Also, every state of a finite system
has an MP representation which thus provides the basis of the powerful
DMRG method. In the thermodynamic limit, second order QPTs occur in
MP ground states accompanied by vanishing energy gaps and diverging
correlation lengths. We have studied the variation of the entanglement
measures $S(i)$ and $S(i,j)$ as a function of $u$ in the ground
state of the spin ladder model with QCPs at $u=0$ and $\infty$.
The major goal of our study is to identify signatures of QPTs, if
any, in the quantum information theoretic measures associated with
entanglement and fidelity. We provide a summary and analysis of our
results below.

Both $S(i)$ and $S(i,j)$ have zero values at $u=\infty$, i.e.,
in the rung singlet phase (Figs. $2$ and $4$) and nonzero values
in the dimerized phase $0<u<\infty$. The entanglement measures can
thus be treated as an order parameter with zero value at the QCP $u=\infty$
and non-zero value in the preceding dimerized phase. In the rung singlet
phase, each rung is described by a spin singlet which is maximally
entangled but the rung is disentangled from the rest of the system.
The EL, $\xi_{E}$, as calculated from $S(i,j)$ diverges as $u\rightarrow\infty$
(Fig. $5$ $(bottom)$) with $\xi_{E}=\frac{\xi_{C}}{2}$, $\xi_{C}$
being the spin-spin correlation length. The entanglement content in
this case vanishes with infinite entanglement range. At the QCP $u=0$,
the entanglement measures have the magnitudes associated with the
AKLT state of a spin-$1$ model. The entanglement measure has a local
minimum at this point, rises to the maximum value at $u=1$ and then
decreases to the global minimum value zero at $u=\infty$. The first
derivatives of $S(i)$ and $S(i,j)$ both fall sharply to zero at
$u=0$ and $u=\infty$. The double derivatives of these quantities
diverge as the QCPs are approached (insets of Figs. $1-4$). The divergence
arises from the structure of the von Neumann entropy involving terms
such as $log_{2}u^{2}$ or $log_{2}\frac{1}{u^{2}}$. A similar type
of divergence occurs in the QPT of a model studied in \cite{key-6}.
We thus find that the entanglement measures $S(i)$ and $S(i,j)$
do develop distinctive features close to the QCPs $u=0$ and $u=\infty$. 

We further looked for signatures of QPTs via the fidelity measure.
Fidelity, i.e., the overlap of ground states for slightly different
Hamiltonian parameters, is expected to drop abruptly at a QCP indicating
a dramatic change in the ground state structure. We plotted $\mathcal{F}(u,\:\delta)=\left\langle u|u+\delta\right\rangle $
with $u$ and $N$ for $\delta=10^{-3}$ and found that the quantity
indeed falls to zero rapidly as the QCPs $u=0$ and $\infty$ are
approached. The quantity $ln|D(u)|$, where $D(u)$ is related to
the second derivative of $F$, also provides a good signature of QPTs.
You $et\, al$ \cite{key-32} has introduced a quantity, the so-called
fidelity susceptibility $\chi_{F}$ which is defined as \begin{equation}
\chi_{F}(u)=lim_{\delta\rightarrow0}\frac{-2\, ln\:\mathcal{F}(u,\:\delta)}{\delta^{2}}\label{25}\end{equation}
One can easily check that $\chi_{F}$ has the same form as $D(u)$.
The finite size scaling hypothesis, which is expected to be valid
in the vicinity of a QCP, leads to the collapse of curves onto a single
scaling function (inset of Fig. $7$) as the QCP $u=\infty$ is approached.
The fidelity measures exhibit similar features in the case of a conventional
QPT. The spin ladder model studied in the paper has spin liquid-type
ground states with none of the phases exhibiting long range magnetic
order in the two-spin correlation functions. The model has three distinct
phases with characteristic quantum order parameters. A characterization
of the transitions between the phases in terms of entanglement and
fidelity measures provide a new perspective on the many body finitely
correlated states and the transitions between them. Quantum information
theoretic measures such as entanglement and fidelity provide a novel
characterization of QPTs occuring in many-body condensed matter systems
\cite{key-33,key-34,key-35}. The present study illustrates this in
the case of a spin ladder model with spin-liquid type ground states.

\section*{ACKNOWLEDGMENT}

A. T. is supported by the Council of Scientific and Industrial Research,
India, under Grant No. 9/15 (306)/ 2004-EMR-I.

\begin{figure}
\includegraphics[%
  scale=0.5]{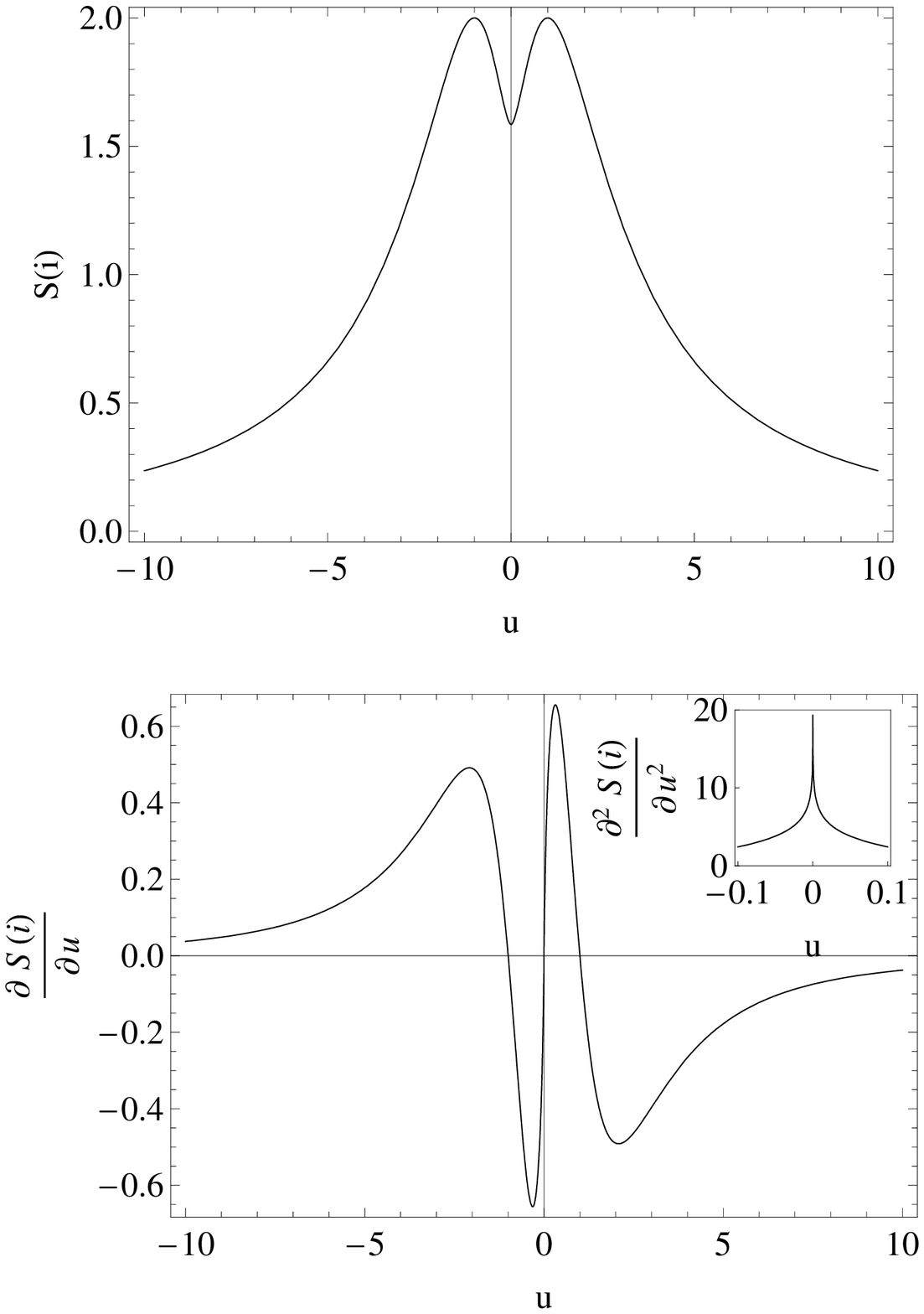}

\textbf{FIG. 1: Plot of $S(i)$ (top) and $\frac{\partial S(i)}{\partial u}$
(bottom) as functions of $u$. The inset (bottom) shows the diverging
behavior of the second derivative of $S(i)$ near $u=0$.}
\end{figure}

\begin{figure}
\includegraphics[%
  scale=0.5]{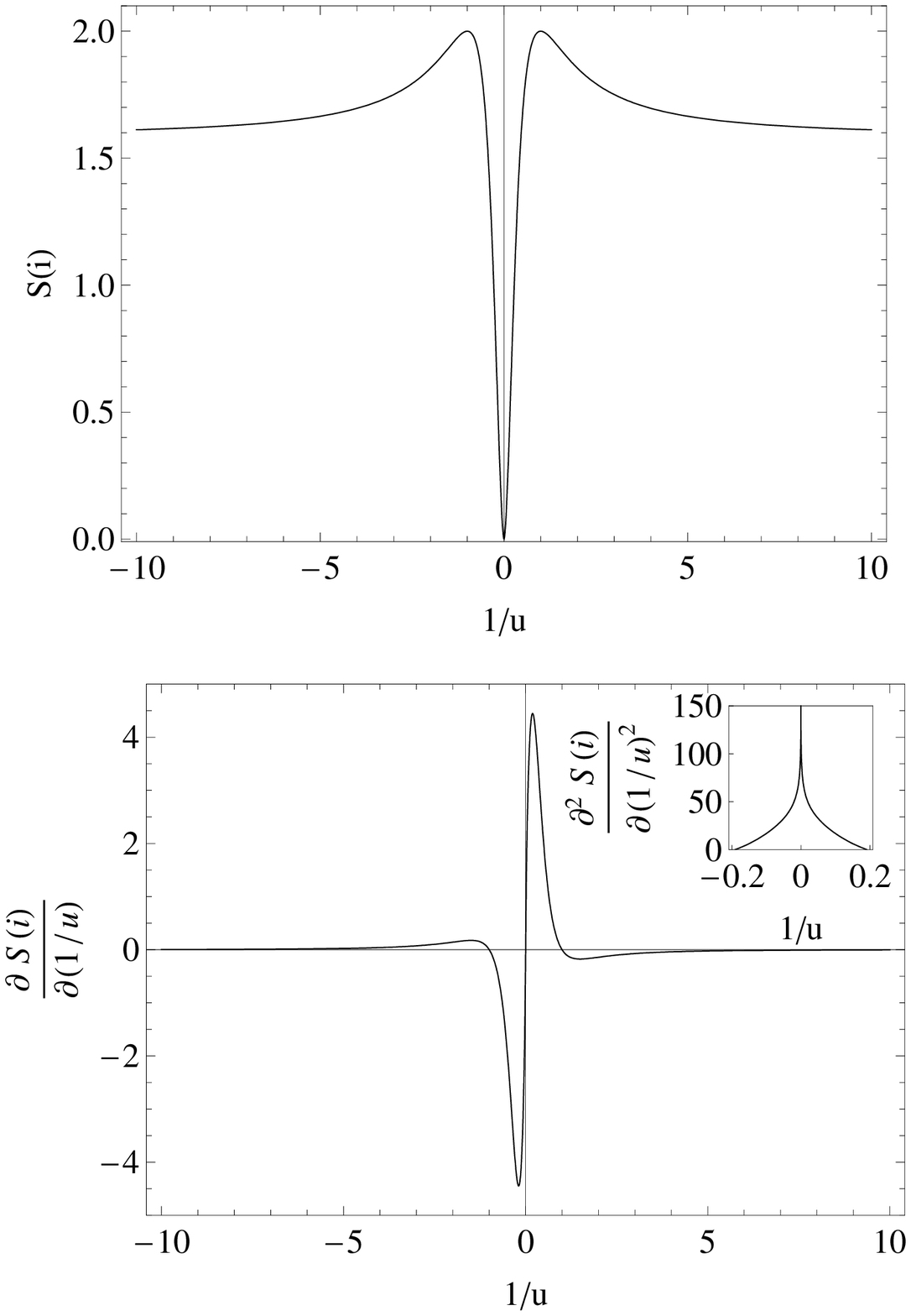}

\textbf{FIG. 2: Plot of $S(i)$ (top) and $\frac{\partial S(i)}{\partial(\frac{1}{u})}$
(bottom) as functions of $\frac{1}{u}$. The inset (bottom) shows
the diverging behavior of the second derivative of $S(i)$ near $u=\infty$.} 
\end{figure}

\begin{figure}
\includegraphics[%
  scale=0.5]{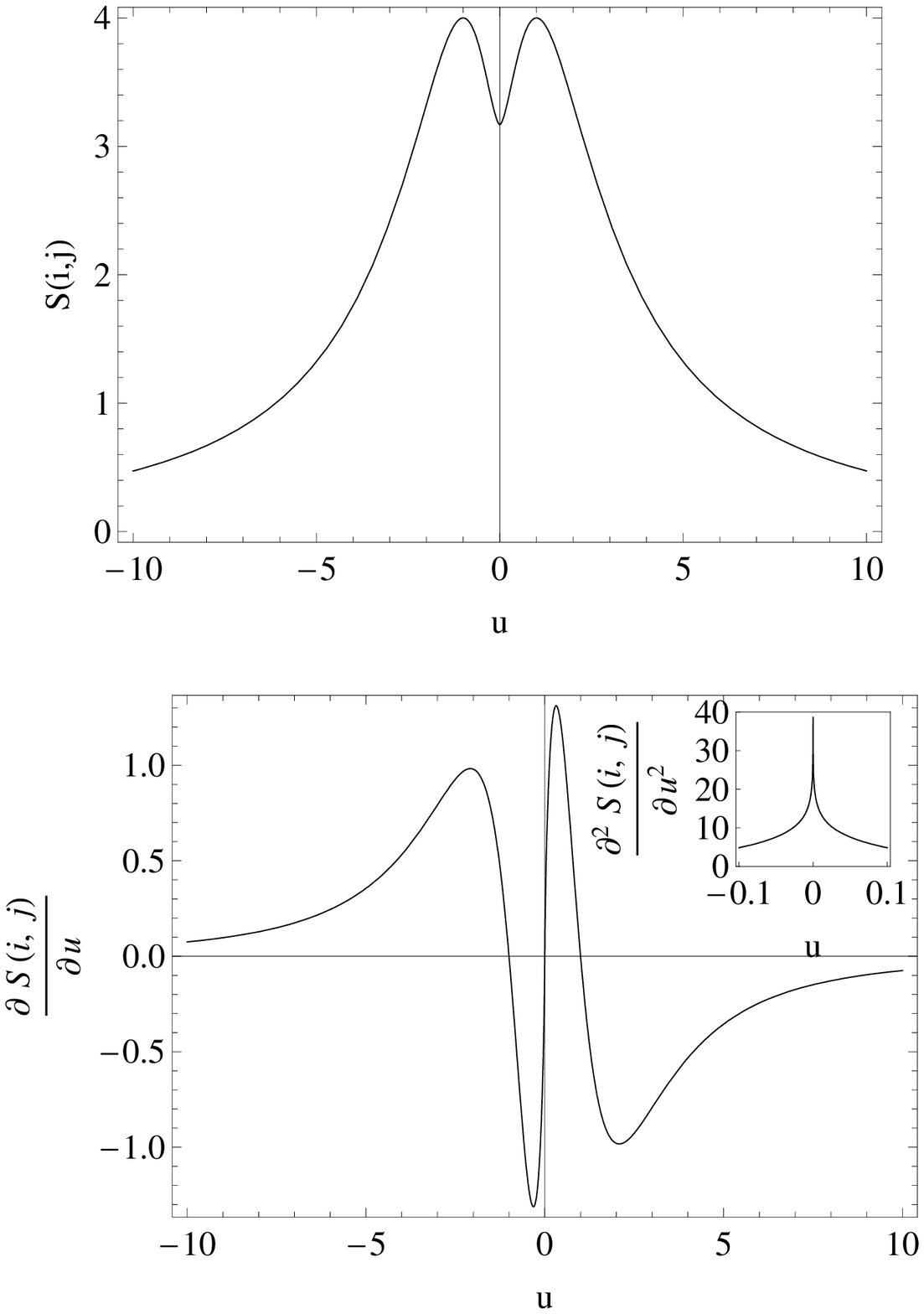}

\textbf{FIG. 3: Plot of $S(i,j)$ (top) and $\frac{\partial S(i,j)}{\partial u}$
(bottom) as functions of $u$ for $n=1000$. The inset (bottom) shows
the diverging behavior of the second derivative of $S(i,j)$ near
$u=0$.} 
\end{figure}

\begin{figure}
\includegraphics[%
  scale=0.5]{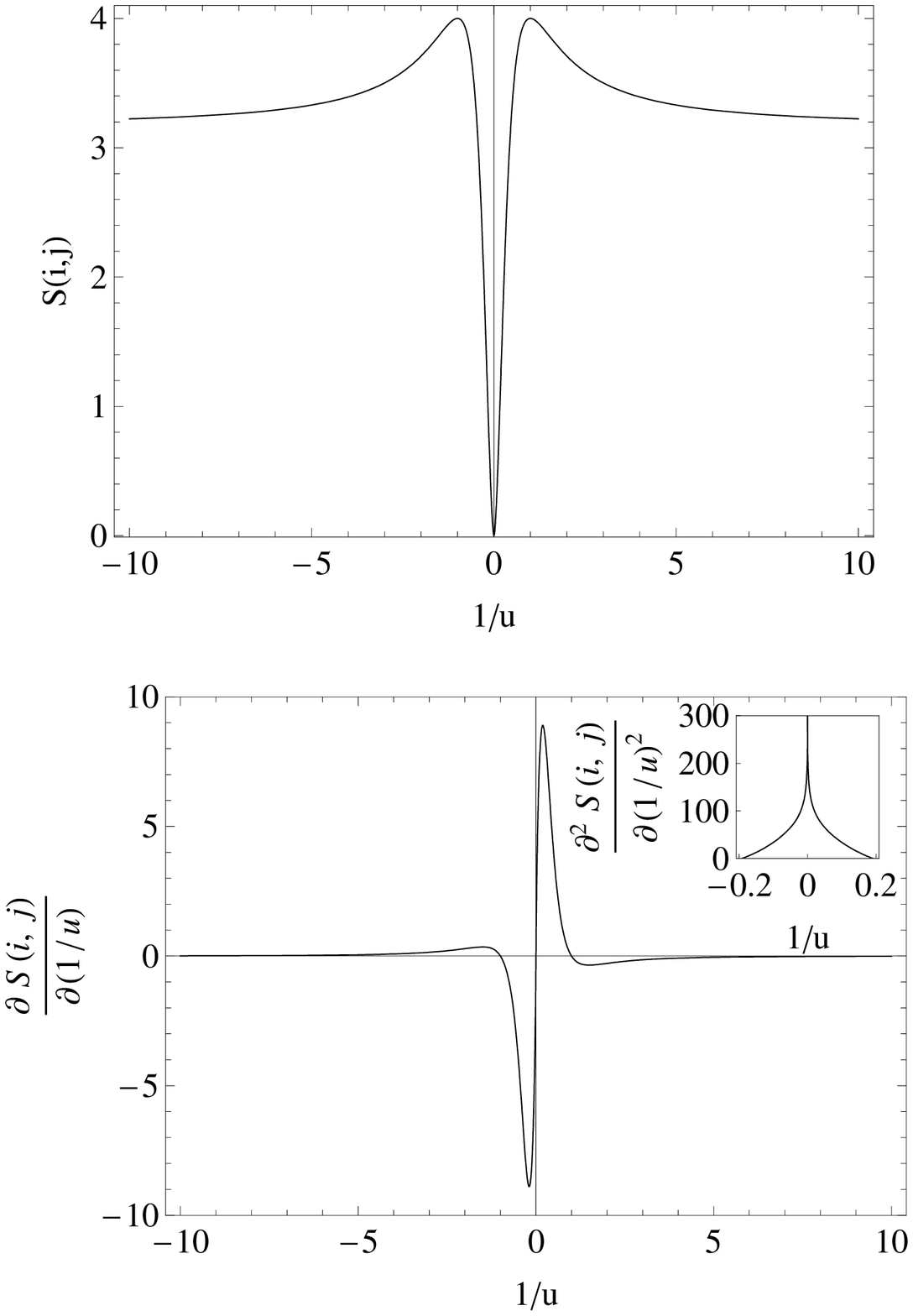}

\textbf{FIG. 4: Plot of $S(i,j)$ (top) and $\frac{\partial S(i,j)}{\partial(\frac{1}{u})}$
(bottom) as functions of $\frac{1}{u}$ for $n=1000$. The inset (bottom)
shows the diverging behavior of the second derivative of $S(i,j)$
near $u=\infty$.} 
\end{figure}

\begin{figure}
\includegraphics[%
  scale=0.8]{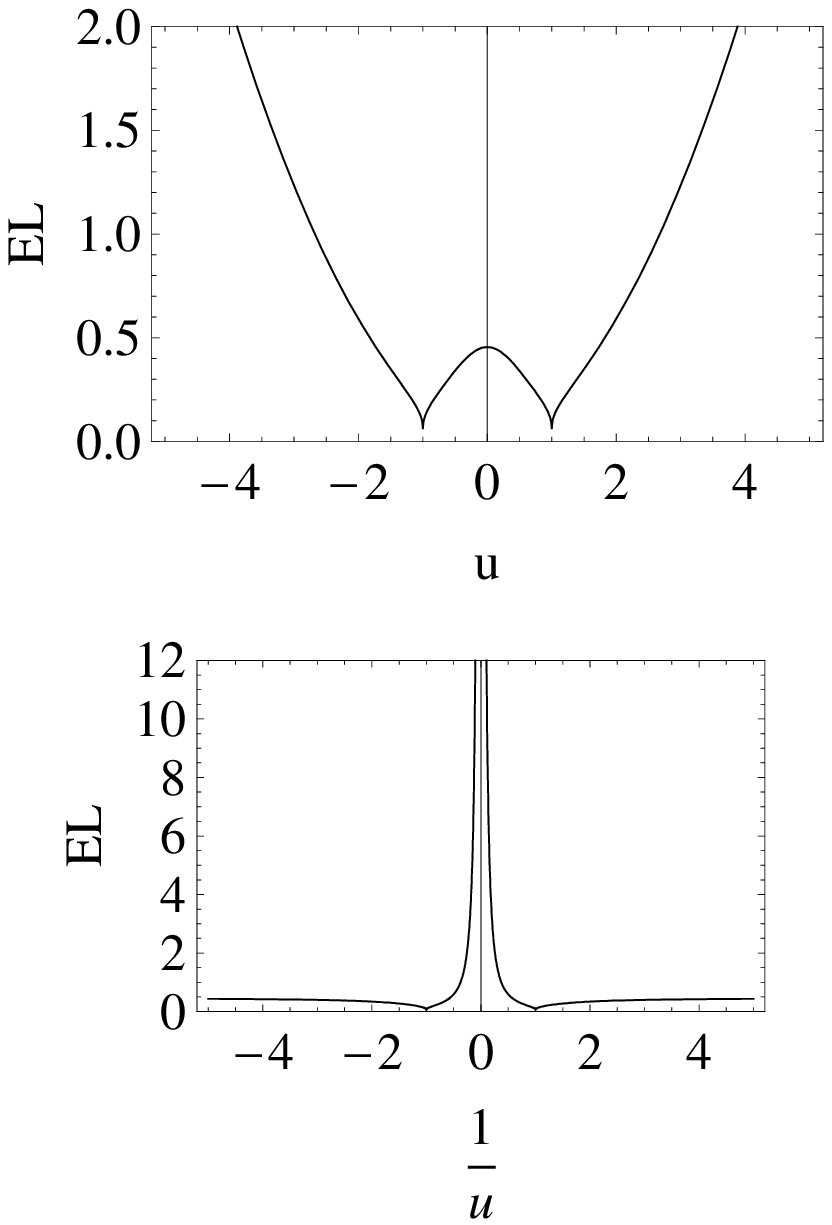}

\textbf{FIG. 5: Plot of EL as a function of $u$ (top) and $\frac{1}{u}$
(bottom) .} 
\end{figure}

\begin{figure}
\includegraphics[%
  scale=0.7]{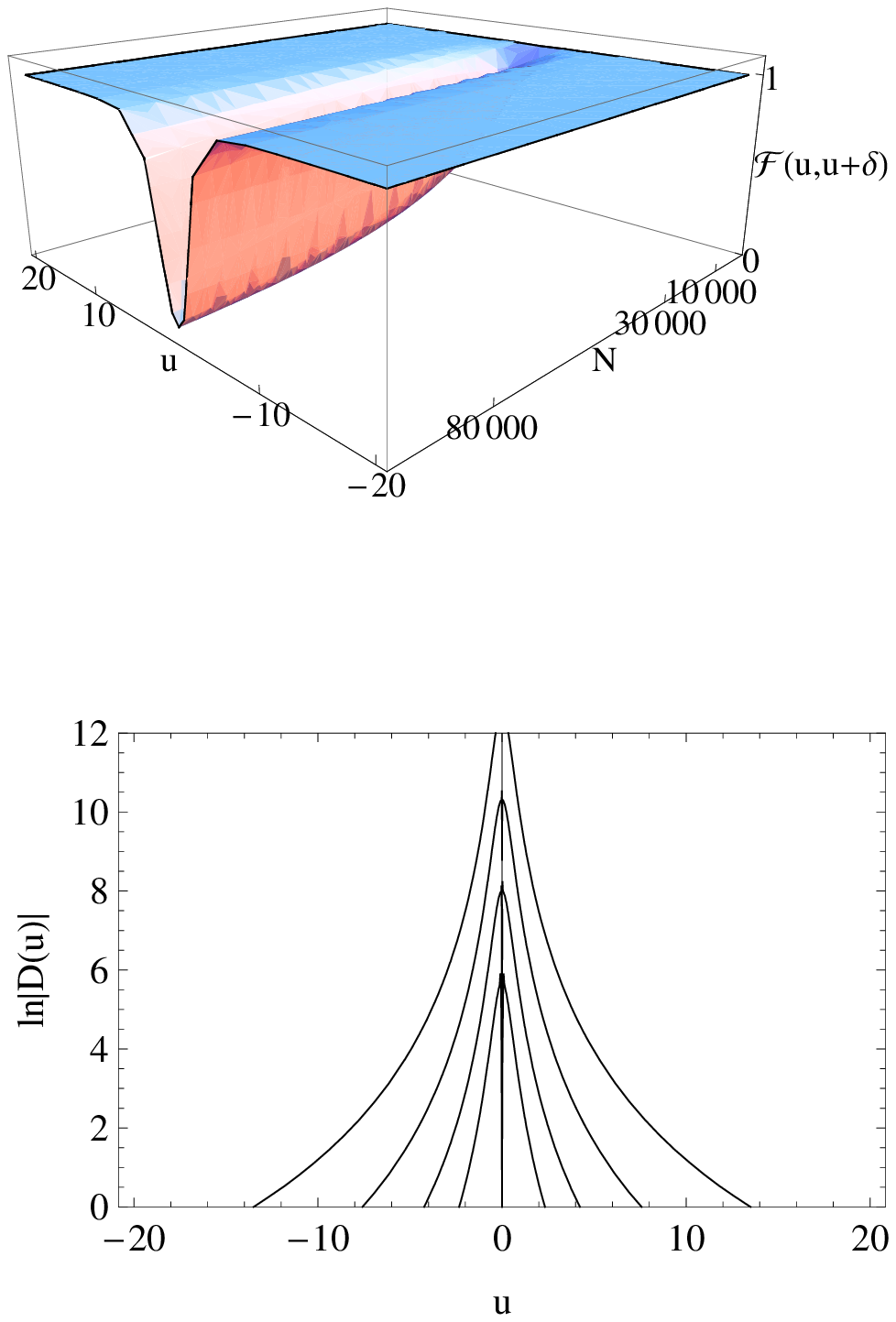}

\textbf{FIG. 6: Plot of $\mathcal{F}(u,\: u+\delta)$ (top) as a function
of $u$ and $N$ for $\delta=.001$ and} $ln|D(u)|$ \textbf{(bottom)
(}$N=10^{3},\,10^{4},\,10^{5}$ \textbf{and} $10^{6}$\textbf{) as
a function of $u$.} 
\end{figure}

\begin{figure}
\includegraphics[%
  scale=0.7]{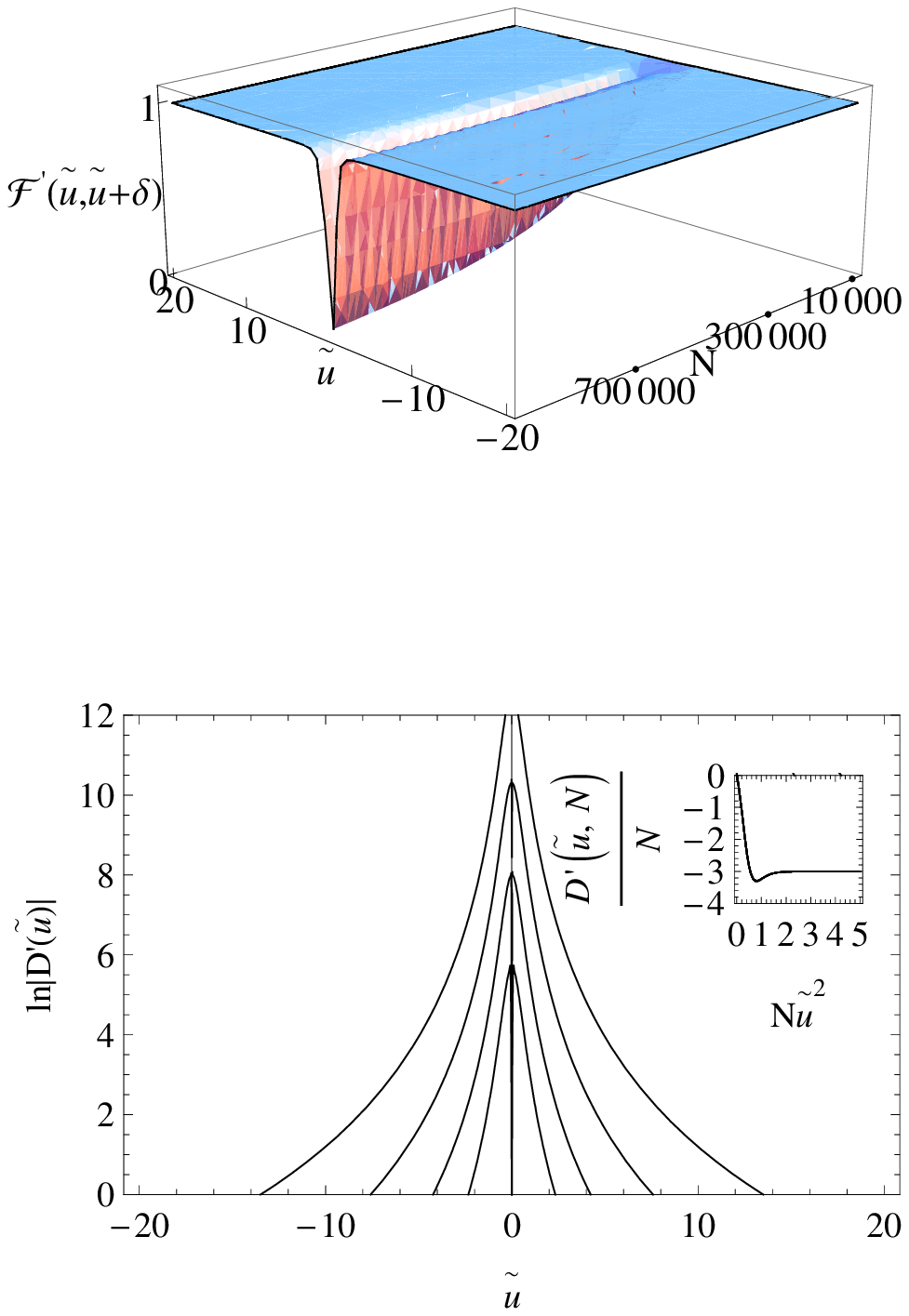}

\textbf{FIG. 7: Plot of} $\mathcal{F}^{'}(\tilde{u},\:\tilde{u}+\delta)$
\textbf{(top) as a function of $\tilde{u}$ and $N$ for $\delta=.001$
and} $ln|D^{'}(\tilde{u})|$ \textbf{(bottom) (}$N=10^{2},\,10^{3},\,10^{4}$
\textbf{and} $10^{5}$\textbf{) as a function of $\tilde{u}$.} \textbf{The
inset shows the data collapse for the rescaled function} $\frac{D^{'}(\tilde{u})}{N}$
\textbf{for same values of} $N$ .
\end{figure}

\end{document}